\documentclass[twocolumn]{revtex4}
\usepackage[cp1251]{inputenc}
\usepackage[english]{babel}
\usepackage{graphicx}
\usepackage{amsmath}
\usepackage{amssymb}
\usepackage{latexsym}

\begin{document}
\title{Quasi-inertial ellipsoidal flows in relativistic hydrodynamics}
\author{Yu.M. Sinyukov$^{1}$, Iu.A. Karpenko$^{1,2}$}
\address{\small \textit{$^{1}$ Bogolyubov Institute for Theoretical
Physics, Kiev 03143, Metrologichna 14b, Ukraine.}}
\address{\small \textit{$^{2}$ National Taras Shevchenko University of Kyiv, Kiev 01033, Volodymyrska 64, Ukraine.}}




\begin{abstract}
We search for non-trivial relativistic solutions of the
hydrodynamic equations with quasi-inertial flows such as in the
Bjorken-like models. The problem is analyzed in general and the
known results are reproduced by a method  proposed. A new class of
3D anisotropic analytic solutions with quasi-inertial property is
found. An ellipsoidal generalization of the spherically symmetric
Hubble flow with constant pressure is proposed as a particular
case. The relativistic expansion of finite systems into vacuum is
also described within this class. A region of applicability and
possible utilization of the new solutions for processes of A+A
collisions is discussed.
\end{abstract}
\maketitle

\section{Introduction}
The equations of the relativistic hydrodynamics have highly
nonlinear nature and, therefore, only a few analytical solutions
are known until now. For the first time one-dimensional, or (1+1),
analytical solution for Landau initial conditions - hot pion gas
in Lorentz contracted thin disk \cite{Landau}, has been developed
by Khalatnikov \cite{Khalatnikov}. The equation of state (EoS) was
chosen as ultrarelativistic one: $p=c_{0}^{2}\varepsilon,
c_{0}^{2}=1/3$. It is noteworthy that according to that solution
the longitudinal flows developed to the end of hydrodynamic
expansion, at freeze-out, are quasi-inertial: $v\approx x_{L}/t$.
Much later, in the papers \cite{Hwa} for the same EoS have been
found the infinite (1+1) boost-invariant solution, for finite
systems the similar approach was developed in Ref.\cite{sin}. The
property of quasi-inertia preserves in these solutions during the
\textit{whole} stage of the evolution. Bjorken \cite{Bjorken}
utilized these solution as the basis of the hydrodynamic model of
ultra-relativistic A+A collisions. \\
\indent The spherically symmetric variant of a such kind of flows
with the Hubble velocity distribution, $v=r/t$, has been
considered in Ref. \cite{Chiu}. Some generalization of these
results was proposed in a case of the Hubble flow for EoS of
massive gas with conserved particle number in Ref.\cite{Csorgo}
and for the cylindrically symmetric
boost-invariant expansion with a constant pressure in \cite{Biro}.\\
\indent All these solutions were used for an analysis of
ultra-relativistic heavy ion collisions. Since the longitudinal
boost-invariance in a fairly wide rapidity region is not observed
even at RHIC, as it was expected, the Hubble-like models are also
used now for a description of the experimental data \cite{flor}.
It is naturally, however, that, unlike to the Hubble type flows,
the velocity gradients should be different in different directions
since there is an initial asymmetry between longitudinal and
transverse directions in central A+A collisions and, in addition,
between in-plane and off-plane transverse ones in non-central
collisions.
   In this letter we make a general analysis of the hydrodynamic equations
for the quasi-inertial flows aiming to find a new class of
analytical solutions with 3D asymmetric relativistic flows.

\section{General analysis}
Let us start from the equations of relativistic hydrodynamics:
\begin{equation}\label{gener}
    \partial_{\nu}T^{\mu\nu}=0,
\end{equation}
where the energy-momentum tensor corresponds to a perfect fluid:
\begin{equation}\label{ideal_fluid}
    T^{\mu\nu}=(\varepsilon+p)u^{\mu}u^{\nu}-p\cdot g^{\mu\nu}
\end{equation}
We can attempt to find a particular class of solutions and
therefore have to make some simplifications of (\ref{gener}). We
do not fix EoS at this stage.
\begin{itemize}
    \item Let us put the condition of quasi-inertiality
    \begin{equation}\label{first eq}
    u^{\nu}\partial_{\nu}u^{\mu}=0
    \end{equation}
    which means that flow is accelerationless in the rest systems of each fluid element; this property holds
    for the known Bjorken (boost-invariant) and  Hubble flows.
    \end{itemize}
    Then, we find that $u^{\mu}[(\varepsilon+p)\partial_{\nu}u^{\nu}+u^{\nu}\partial_{\nu}\varepsilon]
    + [u^{\mu}u^{\nu}\partial_{\nu}p-\partial^{\mu}p]=0$. Contracting this equation with $u_{\mu}$ we get
    \begin{equation}\label{second eq}
    (\varepsilon+p)\partial_{\nu}u^{\nu}+u^{\nu}\partial_{\nu}\varepsilon=0.
    \end{equation}
    Obviously, the remaining equation to satisfy is:
    \begin{equation}\label{third eq}
    u^{\mu}u^{\nu}\partial_{\nu}p-\partial^{\mu}p=0
    \end{equation}

The task is to find solution of the system (\ref{first
eq}),(\ref{second eq}) and (\ref{third eq}). As one can see, the
number of equations exceeds the number of independent variables.
So, the equations must be self-consistent in order to have
nontrivial solutions.

We see that Eq.(\ref{second eq}) can be rewritten in the form:
\begin{equation}\label{second_eq1}
    u^{\mu}\partial_{\mu}\varepsilon = -F(\varepsilon)(\partial_{\nu}u^{\nu})
\end{equation}
where $F(\varepsilon)=\varepsilon+p$, and Eq.(\ref{third eq}) as
the following:
\begin{equation}\label{third_eq1}
    p'(\varepsilon)(u^{\mu}u^{\nu}\partial_{\nu}\varepsilon-\partial^{\mu}\varepsilon)=0,
\end{equation}
supposing  EoS in the form $p=p(\varepsilon)$. If
$p'(\varepsilon)\neq 0$
\begin{equation}
    u^{\mu}F(\varepsilon)(\partial_{\nu}u^{\nu})+\partial^{\mu}\varepsilon=0.
\end{equation}
Normally $F(\varepsilon)\neq 0$, and we can divide the last
equation by $F(\varepsilon)$ and introduce the function
$\Phi(\varepsilon)$ by the definition
$\frac{1}{\varepsilon+p(\varepsilon)}=\Phi'(\varepsilon)$, so that
\begin{equation}\label{energy}
    \partial^{\mu}\Phi(\varepsilon)=-u^{\mu}(\partial_{\nu}u^{\nu})
\end{equation}

Then, the conditions of consistency of equations (\ref{second eq})
and (\ref{third eq}) can be written as:
\begin{equation}\label{cond cons}
    \partial^{\lambda}(u^{\mu}\partial_{\nu}u^{\nu})=\partial^{\mu}(u^{\lambda}\partial_{\nu}u^{\nu})
\end{equation}
In general case, there are 6 independent equations.

   Finally, the relativistic hydrodynamics of
quasi-inertial flows is described by the equations (\ref{first
eq}) and (\ref{cond cons}) for the hydrodynamic velocities
$u^{\mu}$, and the equations (\ref{energy}) for the energy
density: one should use derivative
$\frac{1}{\varepsilon+p(\varepsilon)}=\Phi'(\varepsilon)$ at any
EoS $p=p(\varepsilon)$ to find  function $\varepsilon(x)$. A
serious problem is, however, to find non-trivial solutions for the
field $u^{\mu}(x)$ of hydrodynamic 4-velocities.

\section{Gradient-like velocity ansatz}

One can try to satisfy to Eqs. (\ref{first eq}),(\ref{cond cons})
for velocity profile by a use of gradient-like representation for
it, namely,
\begin{equation}\label{phi}
    u^{\mu}=\partial^{\mu}\phi.
\end{equation}
with condition of normalization, $u^{\mu}u_{\mu}=1$:
\begin{equation}\label{phi1}
    \partial_{\mu}\phi\partial^{\mu}\phi=1
\end{equation}

Then one can check that (\ref{first eq}) is satisfied
automatically, and (\ref{cond cons}) leads to:
\begin{equation}\label{phi2}
    \partial^{\lambda}(\partial^{\mu}\phi\cdot\square\phi)=\partial^{\mu}(\partial^{\lambda}\phi\cdot\square\phi)
\end{equation}
Thus, gradient-like velocity ansatz (\ref{phi}) reduce the problem
to equations (\ref{phi1}),(\ref{phi2}).

One can see that the above equation can be, in particular, reduced
to:
\begin{equation}\label{dalamb}
    \square\phi=F(\phi)
\end{equation}
with any real function $F$ that have to be solved together with
(\ref{phi1}). Note that if $F(\phi)=a+b\phi$ then (\ref{dalamb})
is the linear inhomogeneous partial differential equations (PDE)
and its any solution is a partial solution $\phi_{ih}$ of
inhomogeneous PDE, plus general solution $\phi_{h}$ of
correspondent homogenous PDE (if $b\neq 0$):
\begin{equation}
\begin{array}{l}
\phi_{ih}\sim(t^{2}-\bold{x}^{2}) , b = 0 \\
\phi_{ih}\sim -\frac{a}{b}, b \neq 0
\end{array}
\end{equation}
and
\begin{equation}\label{dalamb_sol2}
    \phi_{h}=\int
d^{4}p\delta (p^2-b)f(p)e^{ipx}
\end{equation}
where $f(p)$ is arbitrary function with properties
$f^{*}(k)=f(-k)$. Then the problem is reduced to a solution of the
nonlinear integral equation (\ref{phi1}) for $f(p)$. If $a=b=0$,
the only potential $\phi=c+c_{0}t+
{\sum}c_{i}x_{i}$ ($i=1,2,3$) with the constrain on the constants $c_{\nu}:c_{0}%
^{2}-{\sum}c_{i}^{2}=1$ is satisfied to these equations. It
describes a relativistic motion of a medium as the whole. It is an
open problem whether there are analytical solutions at $a\neq0$
and/or $b\neq0$.

The known quasi-inertial solutions correspond to $F(\phi)=n/\phi$
in Eq.(\ref{dalamb}). The value $n=1$ generates gradient ansatz
$\phi=\sqrt{t^{2}-z^{2}}$ that gives the (1+1) boost-invariant
Bjorken expansion along axis $z$, $v=z/t$; $n=2$ leads to
$\phi=\sqrt{t^2-x^2-y^2}$, and, correspondingly, to the
two-dimensional (1+2) Hubble-like flow with cylindrical symmetry;
at $n=3$ one can get solution of (\ref{dalamb}) for $\phi$ in the
form $\phi=\tau\doteq\sqrt{t^2-x^{2}-y^{2}-z^{2}}$ describing
spherically symmetric Hubble flow $u^{\mu}=x^{\mu}/\tau$. The
equation (\ref{energy}) has the form $
\partial^{\mu}\Phi(\varepsilon)=nx^{\mu}/\tau^2$ where
number of space coordinates  is equal to $n$. Then the energy
density is described by the following expression
\begin{equation}\label{endens}
\overset{\varepsilon(\tau)}{\underset{\varepsilon(\tau_{0})}{\int}}%
\frac{d\varepsilon}{\varepsilon+p(\varepsilon)}=\ln(\frac{\tau_{0}}{\tau})^{n}.
\end{equation}

\section{Relativistic ellipsoidal solutions}
One more possibility  to satisfy to Eqs. (\ref{third eq}) or
(\ref{third_eq1}) besides of the gradient-like flows is to suppose
a constant pressure in the EoS: ${p^{\prime}(\varepsilon)}=const$.
This possibility was first used in \cite{Biro} as physically
corresponding to a thermodynamic state of the system in the
softest point with the velocity of sound $c_{s}^{2}=0$. Such a
state could be associated with the first order phase transition.
In A+A collisions it corresponds, probably, to transition between
hadron and quark-gluon matter at SPS energies. The solution
proposed in \cite{Biro}  has the cylindrical symmetry in the
transverse plane and the longitudinal boost invariance:
\begin{equation}\label{vb}
    u_{\mu}=\gamma(\frac{t}{\tau},v\frac{x}{r},v\frac{y}{r},\frac{z}{\tau}),
\end{equation}
where $\tau=\sqrt{t^{2}-z^{2}}$, $\gamma=(1-v^{2})^{-1/2}$ and $r$
is transverse radius, $r=\sqrt{x^{2}+y^{2}}$,
\begin{equation}\label{vv}
    v=\frac{\alpha}{1+\alpha\tau}r
\end{equation}
describes axially symmetric transverse flow.

The above solution has, however, a limited region of applicability
since the boost invariance is not expected at SPS energies and can
be used only in a small mid-rapidity interval \cite{AkkMunSin}, it
is not reached even at RHIC energies \cite{Busza}. Most important,
however, is that in non-central collisions there is no axial
symmetry and, therefore, one needs in transversely asymmetric
solutions to describe the elliptic flows in these collisions,
e.g., $v_2$ coefficients. Now we propose a new class of analytic
solutions of the relativistic hydrodynamics for 3D asymmetric
flows.

First we construct the ansatz for normalized 4-velocity:
\begin{equation}
u^{\mu}=\{\frac{t}{\sqrt{t^{2}-\sum a_{i}^{2}(t)x_{i}^{2}}},\frac
{a_{k}(t)x_{k}}{\sqrt{t^{2}-\sum a_{i}^{2}(t)x_{i}^{2}}}\}\label{elliptic sol}%
\end{equation}
where the Latin indexes denote spatial coordinates and $a_{i}$ are
functions of time only. In this case
a set of nonequal $a_{i}$ induces 3D elliptic flow with velocities $v_{i}%
=a_{i}(t)x_{i}/t$: at any time $t$ the absolute value of velocity
is constant,
$\mathbf{v}^{2}=const$, at an ellipsoidal surface $\sum a_{i}^{2}x_{i}%
^{2}=const$. Note that this solution is not gradient-like, so we
follow in the specific way of a further analysis starting from
(\ref{first eq}).

The condition (\ref{first eq}) of accelerationless in this case is
reduced to the ordinary differential equation (ODE) for the
functions $a_{i}(t)$:
\begin{equation}
\frac{da_{i}}{dt}=\frac{a_{i}-a_{i}^{2}}{t}, \label{ak ode}%
\end{equation}
the general solution of which is:
\begin{equation}
a_{i}(t)=\frac{t}{t+T_{i}}, \label{ak}%
\end{equation}
where $T_{i}$ is some set of 3 parameters (integration constants)
having the dimension of time. The different values $T_{1}$,
$T_{2}$ and $T_{3}$ results in anisotropic 3D expansion with the
ellipsoidal flows.

The equation (\ref{third eq}) is satisfied since we assume the
constant pressure profile: $p=p_{0}=const$. The next step is to
find  solution of Eq. (\ref{second eq}) for energy density
$\varepsilon$. Taking into account that
$\partial_{\mu}u^{\mu}=\sum a_{i}/\widetilde{\tau}$, where
\begin{equation}\label{tilde}
    \widetilde{\tau }=\sqrt{t^{2}-\sum a_{i}^{2}x_{i}^{2}},
\end{equation}
one can get
\begin{equation}\label{eq-enthalpy}
(\varepsilon+p_{0})\sum_{i}a_{i}(t)+t\partial_{t}\varepsilon+\sum_{i}a_{i}(t)x^{i}%
\partial_{i}\varepsilon=0.
\end{equation}
General solution of the equation is
\begin{equation}\label{general}
  \varepsilon+p_{0}= \frac{F_{\varepsilon}(\frac {x_{1}}{t+
T_{1}},\frac {x_{2}}{t+T_{2}},\frac {x_{3}}{t+T_{3}})} {(
t+T_{1})(t+T_{2})(t+T_{3})}
\end{equation}
where $F_{\varepsilon}$ is an arbitrary function of its variables.
If one fixes the parameters  $T_{i}$ that define the velocity
profile, then the  function $F_{\varepsilon}$ is completely
determined by the initial conditions for the enthalpy profile,
say, at the initial time $t=0$:
$\varepsilon(t=0,\bold{x})+p_{0}=F_{\varepsilon}(\frac
{x_{1}}{T_{1}},\frac {x_{2}}{T_{2}},\frac
{x_{3}}{T_{3}})/T_{1}T_{2}T_{3}$.

If some value, associated with a quantum number or with particle
number in a case of chemically frozen evolution  are conserved
\cite{AkkMunSin} then one should add the corresponding equation to
the basic ones. Such an equation has the standard form
\cite{Landau2}:
\begin{equation}\label{n_eq}
n\partial_{\nu}u^{\nu}+u^{\mu}\partial_{\mu}n=0
\end{equation}
where $n$ is associated with density of the correspondent
conserved value, e.g., with the baryon or particle densities. A
general structure of this equation is similar to what Eq.
(\ref{eq-enthalpy}) has and, therefore, the solution looks like as
(\ref{general}):
\begin{equation}\label{general1}
n= \frac{F_{n}(\frac {x_{1}}{t+ T_{1}},\frac
{x_{2}}{t+T_{2}},\frac {x_{3}}{t+T_{3}})} {(
t+T_{1})(t+T_{2})(t+T_{3})}
\end{equation}
where the function $F_{n}$ is an arbitrary function of its
arguments and can be fixed by the initial conditions for
(particle) density $n$:
$n(t=0,\bold{x})T_{1}T_{2}T_{3}=F_{n}(\frac {x_{1}}{T_{1}},\frac
{x_{2}}{T_{2}},\frac {x_{3}}{T_{3}})$.

To establish a behavior of other thermodynamic values we use link
between different thermodynamic potentials $\varepsilon=Ts-p+\mu
n$ and utilize the thermodynamic equations based on the free
energy density $f(n,T) = \varepsilon - Ts = \mu n - p$. Since the
volume is fixed (it is unit) the free energy depends on $T$ and
$n$ only, $df=-sdT+\mu dn$, and the chemical potential $\mu =
f_{,n |T=const}$ and the entropy density $s = -f_{,T |n=const}$.

In a case of chemically equilibrated expansion of the
ultrarelativistic gas when the particle number is uncertain and is
defined by the conditions and parameters of the thermodynamic
equilibrium, e.g., by the temperature $T$, the chemical potential
$\mu\equiv 0$ (we suppose here that there is no other conserved
values associated with charges, or the corresponding chemical
potentials are  zero or close to zero). Then $f=-p_{0}=const$,
$df=-sdT=0$ that means the temperature $T=const$ for such a system
and the entropy $s=(\varepsilon(t,\textbf{x})+p_{0})/T$ where
$\varepsilon(t,x)$ is defined by (\ref{general}).

If chemically frozen evolution takes place, the chemical potential
associated with conserved particle number is not zero and
describes the deviation from chemical equilibrium in relativistic
systems. The solution of differential equation $nf_{,n
|T=const}-p_{0}=f(n,T)$ is $f(n,T)=nc(T)-p_{0}$, where $c(T)$ is
some function of the temperature. Then it follows directly from
the thermodynamic identities that
\begin{equation}\label{e-nT}
   \varepsilon(t,\textbf{x}) +p_{0}=
   n(t,\textbf{x})(c(T)-Tc'(T))
\end{equation}
Since the structures of general solutions for $n$ and
$\varepsilon$ are found as (\ref{general}) and (\ref{general1}),
the temperature profile has the form
\begin{equation}\label{T}
    T(t,\textbf{x})=F_{T}(\frac {x_{1}}{t+
T_{1}},\frac {x_{2}}{t+T_{2}},\frac {x_{3}}{t+T_{3}})
\end{equation}
where $F_{T}$ is some function of its arguments that is defined by
the initial conditions for $\varepsilon$ and $n$ and also by EoS
$\varepsilon=\varepsilon(n,T)$. The latter can be fixed by a
choice of the function $c(T)$ in Eq. (\ref{e-nT}). If the initial
enthalpy density profile is proportional to the particle density
profile, $F_{n}(\frac {x_{1}}{T_{1}},\frac {x_{2}}{T_{2}},\frac
{x_{3}}{T_{3}})\sim F_{\varepsilon}(\frac {x_{1}}{T_{1}},\frac
{x_{2}}{T_{2}},\frac {x_{3}}{T_{3}})$, then $T=const$ (and so
$\mu=c(T)=const$) during the system's evolution for any function
$c(T)$ except the linear one: $c(T)=a-bT$ when $T$ is not defined
by the equation (\ref{e-nT}). In the last case
$n=(\varepsilon+p_{0})/a, s=bn$. In another particular case which
corresponds to EoS $\varepsilon+p_{0}=anT$ with $c(T)=-aT\ln(bT)$
one can get:
\begin{equation}\label{Tsimple}
T(t,\textbf{x})=(\varepsilon+p_{0})/(an),\;\;
s(t,\textbf{x})=an(\ln(bT)+1)
\end{equation}
\section{Generalization of the Hubble-like flows}
Let us describe some important particular solutions of the
equations for relativistic ellipsoidal flows. If one defines the
initial conditions on the hypersurface of constant time, say
$t=0$, then $t$ is a natural parameter of the evolution. Such a
representation of the solutions similar to the Bjorken and Hubble
ones with velocity field $v_{i}=a_{i}x_{i}/t$ has property of an
infinite  velocity increase at $x\rightarrow\infty$. A real fluid,
therefore, can occupy only the space-time region where
$|\textbf{v}|<1$, or $\tilde{\tau}^2>0$. To guarantee the
energy-momentum conservation of the system during the evolution,
all thermodynamic densities have to be zero at the boundary of the
physical region, otherwise one should consider the boundary as the
massive shell \cite{sin}. Hence in the standard hydrodynamic
apprach the enthalpy and particle density must be zero at the
surface defined by $|\textbf{v}(t,\textbf{x})|=1$ at any time $t$.
One of a simple form of such a solution (for the case of particle
number conservation) can be obtained from
(\ref{general}),(\ref{general1}) choosing
$F_{\epsilon,n}\sim\exp\left(-b_{\epsilon}^{2}\frac{t^2}{\widetilde{\tau
}^2}\right)$:
\begin{eqnarray}
  \varepsilon(t)+p_{0}=\frac{C_{\varepsilon}}{\prod_{i}(t+T_{i})}\exp\left(-b_{\epsilon}^{2}\frac{t^2}{\widetilde{\tau
}^2}\right),\label{t} \\
n=\frac{C_{n}}{\prod_{i}(t+T_{i})}\exp\left(-b_{n}^{2}\frac{t^2}{\widetilde{\tau
}^2}\right)\label{n}
\end{eqnarray}
where $\tilde{\tau}$ is defined by (\ref{tilde}), and the
constants $C_{\varepsilon}$, $C_{n}$, $b_{\epsilon}$ and $b_{n}$
are determined by the initial conditions as described in the
previous section. As one can see, the enthalpy density tends to
zero when $|\textbf{x}|$ becomes fairly large approaching the
boundary surface defined by $|\textbf{v}(t,\textbf{x})|=1$, in the
other words, when $\tilde{\tau}\rightarrow 0$. Thus the physically
inconsistent situation when massive fluid elements move with the
velocity of light at the surface $\tilde{\tau}=0$ is avoided. Of
course, in such a solution one has to put a constant pressure to
be zero, $p_{0}=0$.

As it follows from an analysis of a behavior of the thermodynamic
values in the previous section, the temperature is constant if
$b_{\varepsilon}=b_{n}$, otherwise one can choose the temperature
approaching zero at the system's boundary, e.g., for EoS which is
linear in temperature, the latter has the form
\begin{align}
T  & =const\text{
\ \ \ \ \ \ \ \ \ \ \ \ \ \ \ \ \ \ \ \ \ \ \ \ \ \ \ \ \ \ \ \ \ \ \ \ }%
b_{\varepsilon}=b_{n}\nonumber\\
& \label{Tdecrease}\\
T  & \sim
e^{-(b_{\varepsilon}^{2}-b_{n}^{2})\frac{t^{2}}{\tilde{\tau}^{2}}}\rightarrow
0, \ \ \ |v(x)|\rightarrow 1 \ \ \ \ \ b_{\varepsilon}>
b_{n}\nonumber
\end{align}
according to (\ref{Tsimple}).

Note that in the region of non-relativistic velocities,
$v^{2}=\sum{\frac{a_{i}^{2}x_{i}^{2}}{t^{2}}}\ll 1$ the space
distributions of the thermodynamical quantities
(\ref{t}),(\ref{n})  has the Gaussian profile:
\begin{eqnarray}\label{nonrel}
\varepsilon+p_{0}  &
\simeq\frac{C_{\varepsilon}}{\prod_{i}(t+T_{i})}e^{-b_{\varepsilon}^{2}\sum{a_{i}^{2}\frac{x_{i}^{2}}{t^{2}}}},\;\label{genHubb}\\
n  & \simeq\frac{C_{n}}{\prod_{i}(t+T_{i})}e^{-b_{n}^{2}\sum{a_{i}%
^{2}\frac{x_{i}^{2}}{t^{2}}}}\nonumber
\end{eqnarray}
The forms of solutions (\ref{nonrel}) are similar to what was
found in Ref. \cite{ellsol} as the elliptic solutions of the
non-relativistic hydrodynamics equations. In this sense the
solution proposed could be considered as the generalization (at
vanishing pressure) of the corresponding non-relativistic
solutions allowing one to describe relativistic expansion of the
finite system into vacuum.

One can note that the case of \textit{equal} flow parameters
$T_{i}=0$ and $b_{\epsilon}=b_{n}=0$ induces formally Hubble-like
velocity profile with the behavior of the density and enthalpy
similar to (\ref{endens}) at n=3, $p=p_{0}$, and with the
substitution $\tau\rightarrow t$.

The direct physical generalization of the Hubble solution for
asymmetric case should be associated with the hypersurfaces of the
pseudo-proper time $\tilde{\tau}$ rather than with time $t$, that
eliminates the problem of infinite velocities:
$v^{2}=\sum{\frac{a_{i}^{2}x_{i}^{2}}{t^{2}}}<1$ at any
hypersurface $\tilde{\tau}^2=const >0$. It can be reached if one
chooses the function $F_{\varepsilon}$ and $F_{n}$ in (\ref
{general}), (\ref{general1}) in the form
\begin{equation} \label{formF}
F\sim\left(\frac{t}{\widetilde{\tau }}\right) ^{3}.
\end{equation}
Then the generalized Hubble solution is
\begin{align}
u^{\mu}  &
=\{\frac{t}{\widetilde{\tau}},\frac{a_{1}x^{1}}{\widetilde
{\tau}},\frac{a_{2}x^{2}}{\widetilde{\tau}},\frac{a_{3}x^{3}}{\widetilde{\tau
}}\},\;\nonumber\\
\varepsilon+p_{0}  &
=C_{\varepsilon}\frac{a_{1}a_{2}a_{3}}{\widetilde{\tau
}^{3}},\;\label{genHubb}\\
n  & =C_{n}\frac{a_{1}a_{2}a_{3}}{\widetilde{\tau}^{3}}\nonumber
\end{align}
where $\widetilde{\tau }=\sqrt{t^{2}-\sum
a_{i}^{2}x_{i}^{2}},\;a_{i}\equiv a_{i}(t)=t/(t+T_{i})$ and
constants are: $C_{\varepsilon
}=T_{1}T_{2}T_{3}(\varepsilon (0,\mathbf{0})+p_{0}),%
\;C_{n}=T_{1}T_{2}T_{3}n(0,\mathbf{0})$. Again, the
proportionality between $\varepsilon$ and $n$ results in the
temperature to be a constant during the evolution. If all
parameters $T_{i}$ are equal to each other, then $a_{i}$ are also
equal and solution (\ref{genHubb}) just corresponds to spherically
symmetric Hubble flow (at constant pressure) and
$\widetilde{\tau}$ is the proper time of fluid element,
$\widetilde{\tau}=\tau$. Note that comparing to the standard
representation of the Hubble solution the origin of a time scale
is shifted, $t\rightarrow t+T_{i}$,  and therefore the singularity
at $t=0$ is absent. Thus, if this solution is applied to a
description of heavy ion collision, $T_{i}$ should be interpreted
as the initial proper time of thermalization and hydrodynamic
expansion to which the origin of a time scale is shifted,
typically $\tau_{0}=T_{i}\simeq1 fm/c$. As to a general case of an
asymmetric expansion, the \textit{minimal} parameter $T_{i}$ can
be considered as the initial time $t$ (at $\textbf{x}=0$) of the
beginning of the hydrodynamic evolution. In analogy with the
Hubble flow the initial conditions in asymmetric case
(\ref{genHubb}) can be ascribed to the hypersurface $\sigma:
\widetilde{\tau}=const$ so that $t_{\sigma}(\textbf{x}=0)=0$.
Note, that such a hypersurface at $|\textbf{x}|\rightarrow\infty$
tends to the hyperbolical hypersurface $\tau = const$ since
$t_{\sigma}(\textbf{x})\rightarrow\infty$ in this limit and so all
$a_{i}\rightarrow 1$.

The boost-invariant (1+1) solutions are also contained in general
ellipsoidal solutions (\ref{general}),(\ref{general1}) for
quasi-inertial flow. To get it one has to choose functions
$F_{\varepsilon}$ and $F_{n}$ in the form (\ref{formF}) with
another power: $3 \rightarrow 1$; it leads to the same form of
solution as (\ref{genHubb}) with replacement $\widetilde{\tau
}^{3}\rightarrow \widetilde{\tau }^{1}$. The next step is to
suppress the transverse flow by setting $T_{1}\rightarrow\infty,
T_{2}\rightarrow\infty$ (as usual, $x_{1}$ and $x_{2}$ denotes
coordinates in the transverse plane and $x_{3}$ is the
longitudinal axis), while the parameter $T_{3}$ is finite. This
limit approach gives us $a_{1}=a_{2}=0$ and $\widetilde{\tau
}\rightarrow\tau=\sqrt{(t+T_{3})^{2}-x_{3}^{2}}$ and results
directly in the Bjorken solution at $T_{3}=0$. Since $T_{i}$ is a
shift of a time scale to the beginning of hydrodynamic expansion,
it is naturally to consider $T_{3}\neq0$ as this was discussed
above for the Hubble-like solution. This value transforms as
$T_{3}\rightarrow T'_{3}=T_{3}/\gamma$ at Lorentz boosts along
axis $x_{3}$.

Note, that if one does not change the power $3\rightarrow 1$ in
(\ref{formF}) and proceed to the limit directly in the equation
(\ref{genHubb}), the particle density (and enthalpy) behavior will
differ from the boost-invariant one as the following:
\begin{equation}
    n\sim \tau ^{-1}\rightarrow n\sim \tau
    ^{-1}\left(1-\frac{x_{3}^{2}}{(t+T_{3})^{2}}\right)^{-1}
\end{equation}
It is also the solution of (1+1) relativistic hydrodynamics at
$p=const$ but it obviously violates the boost-invariance: the
particle and energy densities are not constants at any
hypersurface and their analytic forms are changed in new
coordinates after Lorentz boosts.

It is worthily to emphasize  that the physical solutions with
non-zero constant pressure have a limited region of applicability
in time-like direction: if one wants to continue the solutions to
asymptotically
large times, then $(\epsilon+p_{0})_{t\rightarrow\infty}\approx\frac{C}%
{t^{3}}\rightarrow0$, and this results in non-physical
asymptotical behavior $\epsilon\rightarrow-p_{0}$, unless we set
$p_{0}=0$. Therefore, it is naturally to utilize such kind of
solutions in a region of the first order phase transition,
characterized by the constant temperature and soft EoS,
$c_{s}^{2}=\partial p/\partial\varepsilon \approx 0$, or at the
final stage of the evolution that always corresponds to the
quasi-inertial flows.

\section{Conclusions}
A general analysis of quasi-inertial flows in the relativistic
hydrodynamics is done. The known analytical solutions, like the
Hubble and Bjorken ones, are reproduced from approach developed. A
new class of analytic solutions for 3D relativistic expansion with
anisotropic flows is found.  The ellipsoidal generalization of the
spherically symmetric Hubble flow is considered within this class.
These solutions can also describe the relativistic expansion of
the finite systems into vacuum. They can be utilized for a
description of the matter evolution in central and non-central
ultra-relativistic heavy ion collisions, especially during
deconfinement phase transition and the final stage of evolution of
hadron systems. Also, the solutions can serve as a test for
numerical codes describing 3D asymmetric flows in the relativistic
hydrodynamics.

\section*{Acknowledgments}
We are grateful to S.V. Akkelin, T. Cs\"{o}rg\H{o} and B.
Luk\'{a}cs for their interest in this work and stimulating
discussions. The research described in this publication was made
possible in part by NATO Collaborative Linkage Grant No.
PST.CLG.980086, by Award No. UKP1-2613-KV-04 of the U.S. Civilian
Research $\&$ Development Foundation for the Independent States of
the Former Soviet Union (CRDF) and Ukrainian State Fund of the
Fundamental Researches, Agreement No. F7/209-2004.

    \begin{quote}
    \bibitem{Landau}L.D. Landau, Izv.Acad.Nauk SSSR, s. Fiz. \textbf{17} (1953) 51;
    \bibitem{Khalatnikov} I.M. Khalatnikov, Sov. JETF, \textbf{26} (1954);
    529;
    \bibitem{Hwa} R.C. Hwa, Phys.Rev. \textbf{D10} (1974) 2260; F. Cooper,
    G. Frye, E. Schonberg, Phys.Rev. \textbf{D11} (1975) 192;
    \bibitem{sin} M.I.Gorenstein,Yu.M.Sinyukov,V.I.Zhdanov: Phys.Lett. \textbf{B71} (1977)
    199, and Zh.Eksp. Theor.Fiz.- ZETF (USSR) \textbf{74} (1978), 833;
   \bibitem{Bjorken}  J. D. Bjorken, Phys. Rev. {\bf D27} (1983)
   140;
 \bibitem{Chiu} C.B. Chiu, E.C.G. Sudarshan, Kuo-Hsiang Wang, Phys.Rev. \textbf{D12} (1975) 902;
    \bibitem{Csorgo} T. Cs\"{o}rg\H{o}, Heavy Ion Phys. \textbf{A21} (2004) 73-84;
    \bibitem{Biro} T. S. Biro, Phys.Lett. \textbf{B474} (2000) 21-26, T. S. Biro, Phys.Lett. \textbf{B487} (2000)
    133-139;
   \bibitem{flor}W. Broniowski, W. Florkowski, Nucl. Phys A \textbf{715}, 875c (2003); W. Broniowski, W. Florkowski, B. Hiller, Phys. Rev. C \textbf{68}, 034911 (2003).
    \bibitem{AkkMunSin} S.V. Akkelin, P. Braun-Munzinger, Yu.M. Sinyukov. Nucl.Phys. \textbf{A710} (2002)
    439;
    \bibitem{Busza} W. Busza, Acta Physica Polonica \textbf{B35} (2004) 2873;
    \bibitem{Landau2} L.D. Landau, E.M. Lifshitz "Fluid
    Mechanics", Addison-Westley, Reading, Mass., 1959;
    \bibitem{ellsol}
                S.V. Akkelin, T. Cs\"{o}rg\H{o}, B. Luk\'{a}cs,
                Yu.M. Sinyukov and M. Weiner, Phys. Lett.
                \textbf{B505} (2001) 64.
    \end{quote}

\end{document}